\documentclass[aps,prl,floats,twocolumn,showpacs]{revtex4}

\usepackage{graphicx}
\begin{document}
\title{Direct Measurement of the $g$-Factor of Composite Fermions}
\author{F. Schulze-Wischeler$^{1}$, E. Mariani$^{1,2}$, F. Hohls$^{1}$, and R. J. Haug$^{1}$
  \vspace{1mm}}
\affiliation{$^{1}$Institut f\"ur Festk\"orperphysik, Universit\"at Hannover,
  Appelstra\ss{}e 2, D-30167 Hannover, Germany \\
$^{2}$I. Institut f\"ur Theoretische Physik, Universit\"at Hamburg,
  Jungiusstra\ss{}e 9, D-20355 Hamburg, Germany
\vspace{3mm}}
\date{July 25, 2003}
\begin{abstract}The activation gap $\Delta$ of the fractional quantum Hall states at constant fillings $\nu =2/3$ and $2/5$ has been measured as a function of the perpendicular magnetic field $B$. A linear dependence of $\Delta$ on $B$ is observed while approaching the spin polarization transition. This feature allows a direct measurement of the $g$-factor of composite fermions which appears to be heavily renormalized by interactions and strongly sensitive to the electronic filling factor.
\end{abstract}
\pacs{71.10.Pm, 73.43.Fj, 72.25.Dc}
\maketitle
\newcommand{\omecst}{\omega_{\mathrm{c}}^{*}}
\newcommand{\omec}{\omega_{\mathrm{c}}^{}}
\newcommand{\muB}{\mu_{\tiny{\mathrm{B}}}^{}}

In recent years the role of spin in the fractional quantum Hall
(FQH) regime has been the subject of increasing interest. The ever
improving mobility of the samples allows the observation of the
FQH effect in the low magnetic field regime, where the typical
energy scale associated to the electronic interactions competes
with the Zeeman splitting and thereby mixes different spin
channels. The early observations of a reentrant FQHE at various
fillings \cite{Eisenstein89,Engel92,Du95} in a tilted field
configuration were the first signs of spin polarization
transitions in the ground state. Finally, FQH states with partial
(or vanishing) spin polarization have been directly observed
\cite{Kukushkin99,SmetNature}, in contrast to the fully polarized
nature of the  states in the high field regime.

From a theoretical point of view our understanding of the FQHE has
been deepened by the introduction of Composite Fermions (CF)
\cite{Jain89}, quasiparticles made of one electron bound to an
even number of magnetic flux quanta. Many experiments have
confirmed the extreme versatility of CF in treating the collective
nature of the FQH states in terms of almost-free quasiparticles.
The issue of the effective mass of CF has attracted a lot of
interest and several theoretical predictions have been made on it
\cite{HLR93,JainKamilla,Stern95}.

Much less has been said about the spin-related properties of CF.
CF theory was extended at first time in 1993 to states which are
not fully polarized \cite{Wu93}. The spin-polarization transitions
of the FQH ground state have been the subject of numerical
investigations \cite{Chak84,ParkJain98} prior to their
experimental observations \cite{Kukushkin99}. However, a
simultaneous direct measurement of the parameters entering the CF
mass and $g$-factor is still essentially lacking.

\begin{figure}[ht]  
  \begin{center}
  \resizebox{0.8\linewidth}{!}{{\includegraphics[width=7.0cm]{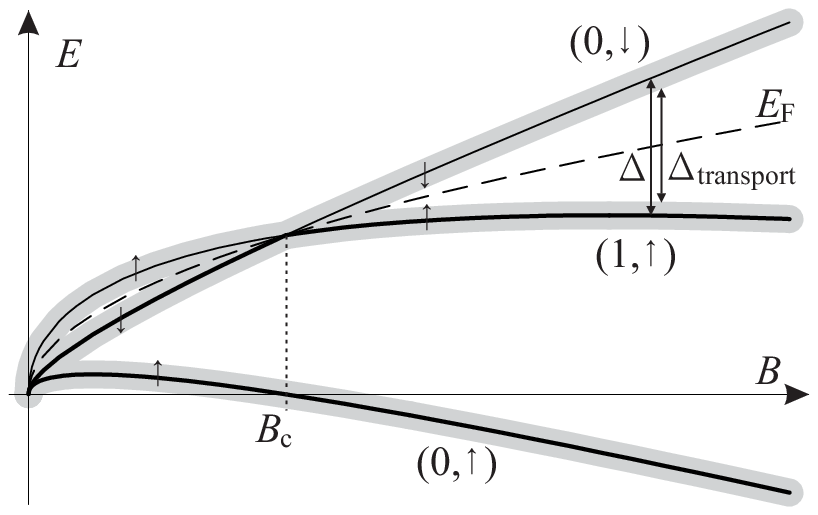}}}
  \end{center}
\caption{The qualitative magnetic field scaling of CF Landau
levels $(n,s)$, with $n$ the Landau level index and $s=\uparrow ,
\downarrow$ the spin. The zero temperature ground states at $\nu
=2/3$ and $2/5$ are obtained by occupying the lowest two CF Landau
levels with spin (thick black lines), with the Fermi energy
(dashed line) lying mid-way between the ground state and the first
excitation (thin line). A quantum phase transition occurs between
a spin-unpolarized and fully-polarized ground state at the
critical magnetic field $B_{\mathrm{c}}^{}$, with a corresponding
activation gap $\Delta$ vanishing linearly as
$|B-B_{\mathrm{c}}^{}|$. The grey regions show the disorder
broadening of the CF Landau levels. Transport measurements are
affected by this disorder and the measured gap
$\Delta_{\mathrm{transport}}$ is systematically smaller than
$\Delta$. However, the broadening depends only weakly on the
magnetic field, and therefore the \emph{slope} of both gaps versus
\emph{B} is \emph{the same}.}
  \label{SCFLLFig}
\end{figure}

A simple model of CF Landau levels with interaction-dependent
cyclotron gap ($\propto \sqrt{B}$) and Zeeman spin-splitting
($\propto B$) is suitable to describe the main structure of the
ground state spin-polarization transitions. The activation gap
approaching the transition is essentially linear in $B$ with a
slope depending \emph{uniquely} on the CF $g$-factor (see
Fig.~\ref{SCFLLFig} and the following text). Previous experimental
analysis of the magnetic fields, where the spin-transition occurs,
produced the filling factor-scaling of the product $m^{*}_{}g$
\cite{Du95,ErosAnn}, with $m^{*}_{}$ the CF mass to be extracted
from the high field activation gap. The information on the CF
$g$-factor was therefore indirect.

In this paper we present an experimental analysis of the
activation gap $\Delta$ at \emph{constant} filling $\nu =2/3$ and
$\nu =2/5$ in the \emph{purely perpendicular field} configuration.
In the CF picture, these two fractions are equivalent, and
correspond to occupying the \emph{two} lowest CF Landau levels.
The sharp linear magnetic field scaling of $\Delta$ while
approaching the spin-polarization transition yields a
\emph{direct} measurement of the CF $g$-factor alone.

Composite Fermions \cite{Jain89} are introduced via the Chern-Simons gauge
transformation on the many-electron wavefunction \cite{Books}. The transformation depends only on the positions of the electrons and is equivalent to attaching an \emph{even} number $\phi $ of flux
quanta $\Phi _{0}=h/e$ to each particle, corresponding to an additional
magnetic field $b(\textbf{r})=\phi \Phi _{0}n_{e}(\textbf{r})$ ($n_{e}(\textbf{r})$
 the local electron density) opposite to the external one. CFs are then subject to an
effective magnetic field $B^{*}(\textbf{r})=B-b(\textbf{r})$ that
vanishes for $\nu \equiv n_{e}\phi _{0}/B=1/\phi$, on the spatial
average (mean field approximation). Near this filling factor the
cancellation is not exact. The residual $B^{*}=B\cdot (1-\phi\nu)$
yields CF Landau levels with an effective cyclotron energy $\hbar
eB^{*}/m_{\mathrm{b}}$ ($m_{\mathrm{b}}$ the electron band mass)
and with a CF filling factor $p=n_{e}\phi _{0}/B^{*}$. The
electronic and CF fillings are related by $\nu =p/(\phi p\pm 1)$
which allows the mean-field mapping of the principal sequence of
the electronic FQH states into integer QHE of CF. In the following
we will consider states around half filling of the lowest Landau
level, thereby choosing $\phi =2$.

In the non-fully spin-polarized case it has been shown that an
independent flux-attachment for the two spin channels produces the
correct principal sequence of FQH states at $\nu=(p_{\uparrow}+p_{
\downarrow})/2[(p_{\uparrow}+p_{\downarrow})\pm 1]$ ($p_{\uparrow/%
\downarrow} $ the numbers of filled spin-up/down CF Landau levels and $p=p_{\uparrow}^{}+p_{\downarrow}^{}$) \cite{Lopez95,Mandal96}. At mean field we have equal cyclotron gaps for the two spin channels.

The mean field assumption has the problem of generating the energy gaps
scaling incorrectly. The dimensional analysis of the spinless case by
Halperin, Lee and Read \cite{HLR93} yields an activation cyclotron gap at
fixed $p$,
\begin{equation}  \label{Delta}
\hbar \omecst\propto \frac{1}{2p\pm 1}
\frac{e^{2}} {\varepsilon l}\,,
\end{equation}
since the Coulomb term $e^{2}/\varepsilon l$ ($\approx 51
\sqrt{B[\mathrm{T}]}\,\mathrm{K}$) is the only relevant energy scale for electrons in the lowest Landau level, with
the dielectric constant $\varepsilon$ ($\approx 12.8$ for GaAs) and $%
l=\left(\hbar/e B\right)^{1/2}$ the magnetic length. The scaling
law (\ref{Delta}) has been confirmed analytically in the large-$p$
limit \cite{Stern95} and by numerical diagonalization of small 2D
systems on a sphere \cite{Morf89}. Equation (\ref{Delta}) can be
obtained by assuming an effective CF mass $m^{*}\propto \sqrt {B}$
\cite{HLR93,JainKamilla} with the activation gap as the smallest
energy needed to excite a CF from the ground state into the first
unoccupied CF Landau level.

Estimates for the magnitudes of the resulting energy gaps have
been obtained without taking into account disorder, finite
thickness of the sample and Landau level-mixing. Thus, in
experiments typically smaller gaps than the theoretically
predicted ones are observed \cite{Boebinger85,Du93}. In order to
discuss the results of our experiments, we introduce the
dimensionless parameter $\alpha$ via $m^{*}_{}/m_{0}^{}=\alpha
\sqrt{B/\mathrm{Tesla}}$ ($m_{0}^{}$ the electron mass in vacuum).
The considerations above suggest the following form of the
effective cyclotron gap at CF-filling $p$
\begin{equation}  \label{homegac}
\hbar\omecst (p,B)=\frac{\hbar eB_{}^{*}}{m_{}^{*}}=\frac{\hbar eB}{m_{}^{*}(2p\pm 1)}\, ,
\end{equation}
consistent with recent numerical investigations \cite{mj01}.

The Chern-Simons transformation does not couple to the electronic
spin degree of freedom. Therefore the Zeeman term can easily be
included and it depends only on $B$. Thus,
\begin{equation}\label{SCFLL}
E_{nps}(B)=\left(n+\frac{1}{2}\right)\hbar\omecst (p,B)
+s\,g\muB B
\end{equation}
are the energies of spin-up/down ($s=\pm 1/2$) CF Landau levels
with $\muB =\hbar e/2m_{0}^{}=0.67\, \mathrm{K}/\mathrm{T}$ (see
Fig.~\ref{SCFLLFig}).

The zero-temperature ground state at a given $B$ is obtained by
occupying the lowest $p$ CF Landau levels. The corresponding spin polarization is $%
\gamma(B)=[p_{\uparrow}(B)-p_{\downarrow}(B)]/p$. Due to the different $B$-scaling of the cyclotron and Zeeman terms in (\ref{SCFLL}), CF Landau levels with opposite spins cross. The transitions
between differently polarized ground states are then given by the crossings between
CF Landau levels at the Fermi energy. For example, the critical
magnetic field $B_{\mathrm{c}}$ at which the transition to the completely
spin-polarized ground state takes place is obtained as the crossing point between
the $(n=0\, , s=\downarrow)$ and the $(n=p-1\, , s=\uparrow)$ CF Landau levels. It can be expressed in terms of the parameter $\alpha$ as
\begin{equation}  \label{Bcritico}
B_{\mathrm{c}}(p)=\left[\frac{2(p-1)}{|g|\,\alpha(2p\pm 1)}\right]^{2}
\end{equation}
in Tesla. From the measurement of $B_{\mathrm{c}}^{}$ we can
therefore extract the product $|g|\,\alpha$. If we linearize the
$B$ dependence of $E_{nps}(B)$ near the crossing we can define the
''slope'' $S_{nps}(B)=\partial _{B}E_{nps}(B)$. It is then easy to
check that
\begin{equation}
\left| S_{np\uparrow }(B_{nn_{}^{\prime
}})-S_{n_{}^{\prime }p\downarrow }(B_{nn_{}^{\prime }})\right| \equiv \partial_{B}^{}\Delta\Big|_{B_{nn_{}^{\prime}}}^{}=\frac{1}{2}\,\left| g\right| \muB,  \label{Slopes}
\end{equation}
with $B_{nn_{}^{\prime }}$ the magnetic field where the two levels
$E_{np\uparrow }$ and $E_{n_{}^{\prime }p\downarrow }$ cross. The
energy gap $\Delta$ approaching the transition is therefore linear
and its slope (\ref{Slopes}) \emph{uniquely} depends on the CF
$g$-factor. Moreover, the relative slopes of the two CF Landau
levels at the crossing is the same for all the possible crossings
at a given filling factor. Thus, a measurement of the linear gap
while approaching the spin-transition \emph{directly} yields the
CF $g$-factor and the value of the critical field
$B_{\mathrm{c}}^{}$ finally determines the effective mass
parameter $\alpha$.

The model above neglects CF-CF interactions which become relevant
\emph{very close} to the spin-polarization transition. They are
responsible for the fascinating partly-polarized state occurring
in the middle of the transition \cite{Kukushkin99,Murthy00}. The
typical energy scale involved in the formation of this state is
$\delta\cong 0.2\, \mathrm{K}$, so that the gap-linearization is
suitable for energies typically larger than $\delta$, as we have
in our measurements.

In what follows we will concentrate on the activation gap
measurement for fixed $\nu =2/3$ and $2/5$ (i.e. $p~=~2$) and
deduce the CF $g$-factor in consistence with the depicted model.

Many activation measurements around the spin-polarization
transitions were made in a tilted field geometry
\cite{Eisenstein89,Engel92,Du95}. The consequent in-plane field
couples to the finite thickness of the 2DES, modifying the
effective 2D Coulomb interaction \cite{Haug87} and thereby
affecting the mass parameter $\alpha$. In our measurement the
magnetic field is \emph{purely perpendicular}. We believe this
geometry allows a cleaner determination of the CF parameters.

The two-dimensional electron system used in our experiments was
realized in an AlGaAs/GaAs-heterostructure and the carrier density
was modulated by illumination. The base carrier density and
mobility are $n_e= 0.89\cdot 10^{15}\, {\rm m}^{-2}$ and $\mu_e =
102\, {\rm m^2/Vs}$. With maximal illumination $n_e= 1.50\cdot
10^{15}\,{\rm m}^{-2}$ and $\mu_e = 193\,{\rm m^2/Vs}$ are
achieved. The sample was patterned into a long meandering bar with
length $l = 10$ mm and width $w = 90\,\mu$m. The large aspect
ratio $l/w=111$ allows to measure small resistance changes at the
resistance minima. Contacts to the bar were realized by standard
Au/Ge/Ni alloy annealing, yielding negligible contact resistances
$R_{\mathrm{c}}^{} < 10\,\Omega$.

\begin{figure}[ht]  
  \begin{center}
  \resizebox{0.8\linewidth}{!}{\rotatebox{270}{\includegraphics{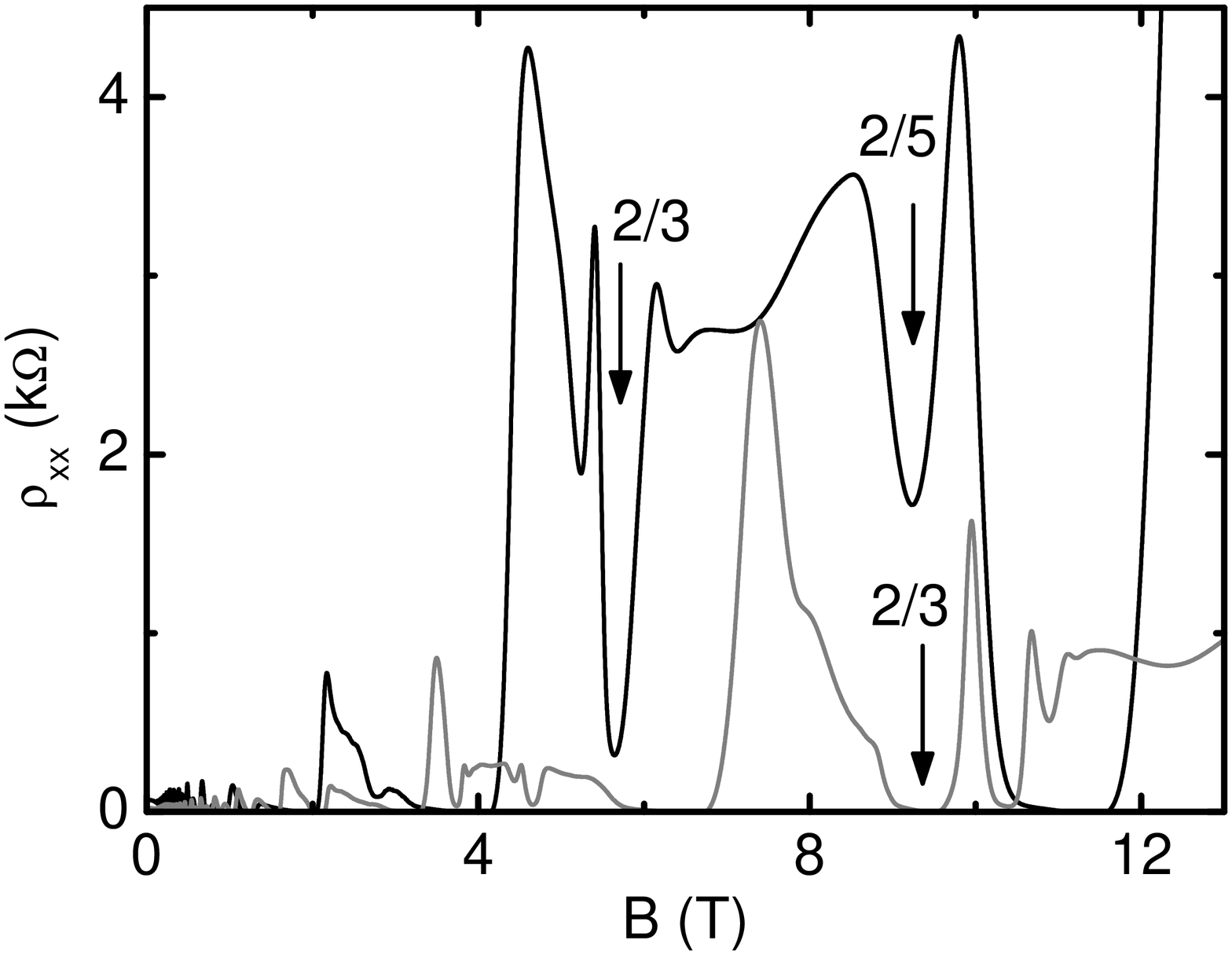}}}
  \end{center}
  \caption{Resistivity $\rho_{\mathrm{xx}}^{}(B)$ without illumination
  ($n_e= 0.89\cdot 10^{15} {\rm m}^{-2}$, black line) and for maximal
  illumination ($n_e= 1.50\cdot 10^{15} {\rm m}^{-2}$, grey line). Arrows mark the
  positions of the filling factors $\nu = 2/3$ and $\nu=2/5$ at which
  activation measurements are performed.
    }
  \label{SdH}
\end{figure}

The sample is mounted onto the cold finger of a He$^3$/He$^4$
dilution refrigerator and is placed at the center of a
superconducting solenoid capable of producing fields up to
$B=13\,$T. An infrared LED allows to change the carrier density
$n_e$ using the persistent photoconductivity. From the two-point
resistance $R_{\mathrm{2p}}^{}$ we calculate the longitudinal
resistivity $\rho_{\mathrm{xx}}^{} =
\left(R_{\mathrm{2p}}^{}-R_{\mathrm{H}}^{}\right)\cdot w/l$ with
$1/R_{\mathrm{H}}^{} = \nu e^2/h$ at integer and fractional
filling factors $\nu$. Starting with the unilluminated sample with
lowest carrier density, we change $n_e$ stepwise by illumination
at zero magnetic field. Monitoring the resistance during the
illumination allows a good control of the increase in $n_e$. For
different $n_{e}^{}$ we measure the resistivity
$\rho_{\mathrm{xx}}^{}$ as function of the magnetic field $B$,
which allows to determine carrier density and mobility. Figure
\ref{SdH} shows $\rho_{\mathrm{xx}}^{}$ for the smallest and
largest density.

\begin{figure} [ht] 
  \begin{center}
  \resizebox{0.95\linewidth}{!}{\rotatebox{270}{\includegraphics{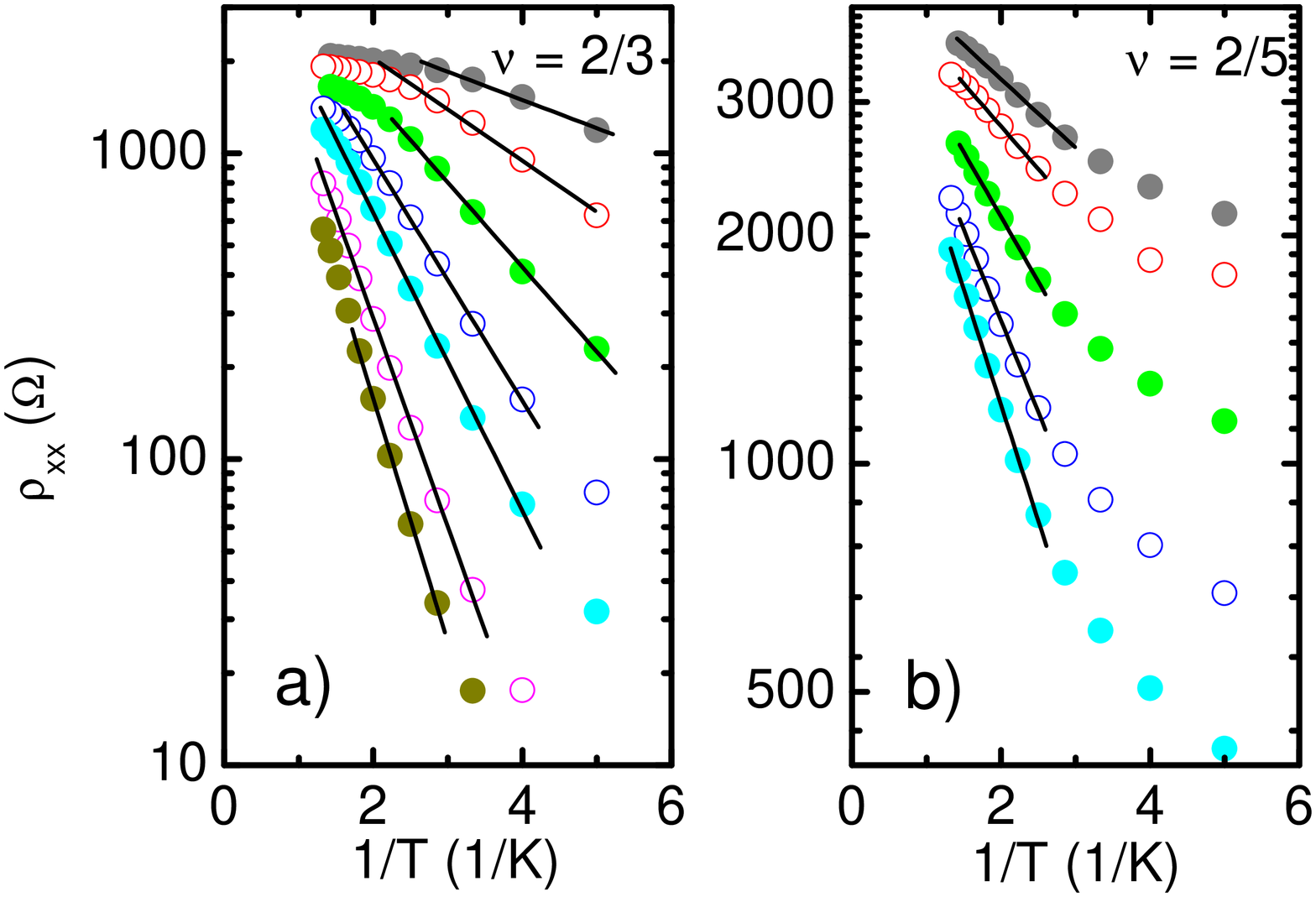}}}
  \end{center}
  \caption{Temperature dependence of the resistivity $\rho_{\mathrm{xx}}^{}(T)$ for
    filling factors $\nu = 2/3$ and $\nu=2/5$ at different densities resp.\
    different magnetic fields shown as Arrhenius plot. The lines show fits of
    activated transport $\rho_{\mathrm{xx}}^{}(T)\propto \exp\left(-\Delta/2T\right)$ to our
    data. a) Filling factor $\nu = 2/3$ at magnetic fields ranging from
    $B= 5.64\,$T (topmost) to $B=9.08\,$T (bottom curve). b) Filling factor
    $\nu = 2/5$ at magnetic fields ranging from $B= 9.23\,$T (topmost)
    to $B=12.55\,$T (bottom curve).
    }
  \label{activation}
\end{figure}

\begin{figure} [ht] 
  \begin{center}
  \resizebox{0.75\linewidth}{!}{\rotatebox{270}{\includegraphics{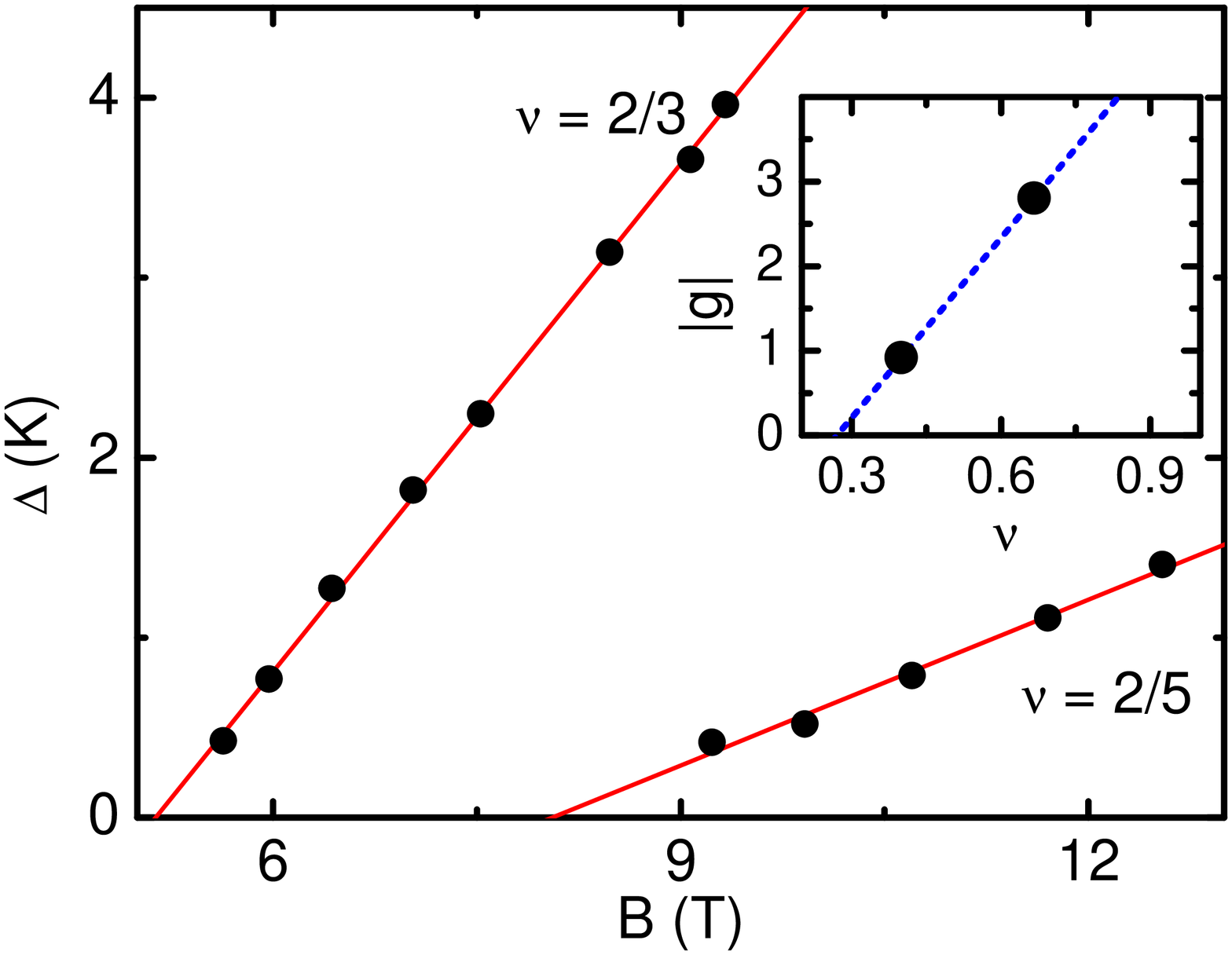}}}
  \end{center}
  \caption{Activation gaps $\Delta(B)$ at filling factors $\nu = 2/3$ and
  $\nu=2/5$ as a function of the magnetic field. $\Delta$ is determined from the
  fits displayed in Fig.~\ref{activation}. The activation gaps are nicely fitted
  by a linear dependence as shown by the solid lines. The slopes of these lines
  are directly related to the CF $g$-factor by $0.5 |g| \muB$. {\it Inset}: $g$-factor
  extracted from the linear dependence of the activation energy $\Delta$. The two
  values of $|g|=0.92$ and $2.80$ are obtained at $\nu =2/5$ and $\nu =2/3$. The dashed
  line shows a linear interpolation.
    }
  \label{gfactor}
\end{figure}

For each density we measured the temperature dependence of the
resistivity $\rho_{\mathrm{xx}}^{}(T)$ while fixing the magnetic
field corresponding to filling factors $\nu =2/3$ and $2/5$. The
result is shown in Fig.~\ref{activation}. As expected for these
fractional filling factors we observe activated transport
$\rho_{\mathrm{xx}}^{} \propto \exp\left(-\Delta/2T\right)$ with
$\Delta$ the activation gap. From the fits to this equation, shown
as solid lines in Fig.~\ref{activation}, we extract $\Delta$ at
$\nu = 2/3$ and $\nu = 2/5$ for different densities resp.\
magnetic fields.
The dependence of the activation gap $\Delta(B)$ on the magnetic
field is shown in Fig.~\ref{gfactor}.

A remarkably linear behaviour is observed over a large magnetic
field range, in agreement with the theoretical expectation. Here
we want to point out that activation gaps measured by transport
are typically reduced by disorder. However, as this effect is
expected to be only weakly depending on $B$ the measured
$\Delta_{\mathrm{transport}}$ (see Fig.~\ref{SCFLLFig}) differs
from $\Delta$ only by a constant. Therefore the \emph{slope}
$\partial_{B}^{}\Delta (B)$ is essentially \emph{the same}. From
the slope $\partial_{B}^{}\Delta (B)$ in formula (\ref{Slopes}) we
extract the CF $g$-factor. Our measurement yields
$|g_{2/5}^{}|=0.92$ and $|g_{2/3}^{}|=2.80$.

Two important considerations come out. First, the value of $g$
indicates a strong renormalization due to interactions (the bulk
$g$-factor for GaAs being $-0.44$). Second, the CF $g$-factor
depends \emph{strongly} on the \emph{electronic} filling factor.
In fact, although both the two fractions $\nu =2/3$ and $2/5$ are
mapped into the same CF-filling $p=2$ (symmetrically with respect
to $\nu =1/2$) their $g$-factors differ by more than a factor 3.

The determination of the slope $\partial_{B}^{}\Delta (B)$ yields
a precious experimental information since it is essentially
unaffected by disorder-broadening of the CF Landau levels. On the
contrary, the activation measurement is not extremely accurate for
the determination of the critical field $B_{\mathrm{c}}^{}$, since
disorder tends to "close the gap" before the spin-polarization
transition occurs.

Since our CF form around $\nu =1/2$ it is interesting to estimate
$g_{1/2}^{}$. From a linear interpolation of our measurement
depicted in the inset of Fig.~\ref{gfactor} we obtain
$|g_{1/2}^{}|=1.65$, in excellent agreement with the data by
Kukushkin \cite{Kukushkin99} which yield $|g_{1/2}^{}|=1.6$ when
using their $\alpha \approx 0.2$ (see \cite{ErosAnn} for details).
In contrast, NMR experiments in \emph{tilted fields} found
$|g_{1/2}^{}|\approx 0.39$ \cite{Freytag02}. The $g$-factor
measured in the NMR experiments seems to be roughly consistent
with the bare $g$-factor of 2D electrons in GaAs, whereas our
measured $g$-value seem to be related to the exchange enhanced
electronic $g$-factor. For the exchange enhanced electronic
$g$-factor a linear scaling with filling factor is expected from
theory \cite{AFS}.

Using this linearization of the $g$ vs.\ $\nu$ dependence, we
could extrapolate to get $|g_{\nu =1}^{}|=5.2$, which would be
roughly consistent with the experimental results of the exchange
enhanced electronic $g$-factor at $\nu$ =1 \cite{Nicholas88}. In
the small $\nu$ regime ($\nu <1/3$) we should be more careful
since the range of the four-flux CF ($\phi =4$) would be explored
with the relative (probably different) $g$-factor.

In conclusion, we performed a \emph{direct} measurement of the CF
$g$-factor for $\nu =2/5$ and $2/3$. Although the two fractions
are \emph{equivalent} in the CF picture, their $g$-factor is
significantly different, showing interaction-renormalization and a
strong dependence on the \emph{electronic} filling factor $\nu$.

We acknowledge discussions with Bernhard Kramer. The sample was
grown by D. Reuter and A. D. Wieck at the University of Bochum. We
acknowledge financial support by BMBF, TMR and DFG.


\end{document}